\documentclass[aps, prd, twocolumn, superscriptaddress, nofootinbib]{revtex4-1}
\usepackage{graphicx}
\usepackage[caption=false]{subfig}
\usepackage{amssymb}
\usepackage{amsmath}
\usepackage{braket}
\usepackage{listings}
\usepackage{subfig}
\usepackage{cases}
\usepackage{comment}
\usepackage[colorlinks=true
,urlcolor=blue
,anchorcolor=blue
,citecolor=blue
,filecolor=blue
,linkcolor=blue
,menucolor=blue
,pagecolor=blue
,linktocpage=true
,pdfproducer=medialab
,pdfa=true
]{hyperref}
\newcommand{\dif}[2]{\frac{\mathrm{d} #1}{\mathrm{d} #2}}
\newcommand{\pdif}[2]{\frac{\partial #1}{\partial #2}}
\newcommand{\dd}{\mathrm{d}}
\newcommand{\ee}{\mathrm{e}}
\def\slashchar#1{\setbox0=\hbox{$#1$} 
\dimen0=\wd0 
\setbox1=\hbox{/} \dimen1=\wd1 
\ifdim\dimen0>\dimen1 
\rlap{\hbox to \dimen0{\hfil/\hfil}} 
#1 
\else 
\rlap{\hbox to \dimen1{\hfil$#1$\hfil}} 
/ 
\fi}

\newcommand{\ltsim}{\protect\raisebox{-0.5ex}{$\:\stackrel{\textstyle <}
	{\sim}\:$}}
\newcommand{\gtsim}{\protect\raisebox{-0.5ex}{$\:\stackrel{\textstyle >}
	{\sim}\:$}}

\begin{document}
\title{Primordial black holes as biased tracers}
\date{\today}
\author{Yuichiro Tada}
\email{yuichiro.tada@ipmu.jp}
\affiliation{Kavli Institute for the Physics and Mathematics of the Universe (WPI), The University of Tokyo, Kashiwa, Chiba 277-8583, Japan}
\affiliation{Department of Physics, the University of Tokyo, Bunkyo-ku
113-0033, Japan}
\author{Shuichiro Yokoyama}
\email{shuichiro@rikkyo.ac.jp}
\affiliation{Department of Physics, Rikkyo University,\\ 3-34-1 Nishi-Ikebukuro, Toshima, Tokyo
171-8501, Japan}
\preprint{IPMU 15-0014}
\preprint{RUP-15-3}

\begin{abstract}

Primordial black holes (PBHs) are theoretical black holes which can be formed during the radiation dominant era  
through the gravitational collapse of radiational overdensities. 
It has been well known that in the context of the structure formation in our Universe such collapsed objects, e.g., halos/galaxies,
could be considered as bias tracers of underlying matter fluctuations and the halo/galaxy bias has been studied well.
Applying such a biased tracer picture to PBH,
we investigate the large scale clustering of PBH and
obtain an almost mass-independent constraint to the scenario that the dark matter (DM) consists of PBH. 
We focus on the case where the statistics of the primordial curvature perturbations is almost Gaussian, but with small local-type non-Gaussianity.
If PBHs account for the DM abundance, such a large scale clustering of PBHs 
behaves as nothing but the matter isocurvature perturbation which is strictly constrained by the observations of cosmic microwave backgrounds (CMBs).
From this constraint, we show that, in the case where a certain single field causes both CMB temperature perturbations and PBH formation,
the PBH-DM scenario is excluded even with quite small local-type non-Gaussianity, $|f_\mathrm{NL}|\sim\mathcal{O}(0.01)$.

\end{abstract}

\maketitle

\section{Introduction}
The origin of the dark matter (DM) is one of the most earnest questions for scientists.
A lot of candidates have been proposed, but it still remains an open question.
A primordial black hole (PBH)~\cite{Carr:1975qj} is one of the DM candidates, which has been studied for a long time.
PBHs are black holes which are theoretically suggested to be formed in the early universe,
and distinguished from ordinary black holes which are believed to be formed from the collapse of stars.  

The constraint on the abundance of PBHs has been basically obtained from the non-detection of them 
(see Ref. \cite{Carr:2009jm}),
and recently, by the Kepler telescope, 
all of the mass windows for PBHs to be a dominant component of DMs would be closed and the possibility of PBH-DM scenario
seems to be completely excluded 
if PBHs have a monochromatic mass function~\cite{Griest:2013aaa,Capela:2013yf}.
However, there remain theoretical uncertainty of constraints, the possibility such that the mass region of PBHs
broadens widely, and so on.

In this paper, 
we consider the large scale clustering behavior of PBHs and 
discuss the possibility of PBH-DM scenario in a way independent of the PBH mass.
Here, we assume the case that PBHs are formed from the gravitational collapse of the radiational overdense regions, though
there are several mechanisms for the PBH formation like a collapse of cosmic strings~\cite{Polnarev:1988dh} and domain walls~\cite{Rubin:2000dq}, 
or bubble collisions~\cite{Hawking:1982ga}. 

In the context of the structure formation,
the large scale clustering of such collapsed objects as halos/galaxies, has been studied well.
For example, Bardeen \emph{et al.}~\cite{Bardeen:1985tr} showed that the collapsed objects tend to 
be formed intensively in widely dense regions, in the peak-background split picture in which source density perturbations are
divided into short- (peak) and long-wavelength (background) modes. 
In the case where the statistics of the initial primordial perturbations is Gaussian,
the large scale clustering of the collapsed objects could be related to the distribution of the underlying matter density field by introducing 
a constant \emph{scale-independent bias} parameter.
On the other hand,
due to the existence of the non-Gaussian feature in the statistics of the initial primordial perturbations,
the bias parameter tends to have a scale-dependent component, which is known as \emph{scale-dependent bias},
and this has been considered to be a powerful tool to hunt the primordial non-Gaussianity and extensively studied (see e.g., Ref.~\cite{Dalal:2007cu}).

With the object that PBHs are also collapsed objects,
Chisholm~\cite{Chisholm:2005vm} considered the scale-independent bias of PBHs and concluded that 
though the bias factor is large in itself, 
the density perturbations of PBHs are hardly produced on much larger scales like observable scales of cosmic microwave backgrounds (CMB) temperature anisotropies.
Recent several papers~\cite{Young:2014ana,Nakama:2014fra} also agree with this result.
This is simply because that PBHs are caused by the collapse of horizon scale overdensity, which are hardly affected by the super-horizon density fluctuations
that has little causality with physics inside the horizon.

In this paper, 
based on these previous works, we consider the case where the statistics of the primordial curvature perturbation
is almost Gaussian, but with small non-Gaussian component, and investigate the large scale clustering of the PBHs.
As a non-Gaussian component, we consider the so-called local-type non-Gaussianity, which is known to give a strong scale-dependence
in the halo bias.
We find that if PBH is a dominant component of DM, the large scale clustering of the PBH can be detected as 
the matter isocurvature perturbations and excluded by CMB observations even in the case with small non-Gaussianity.

This paper is organized as follows. In section~\ref{Single source case}, we briefly review the bias effect and calculate
the matter isocurvature perturbations from biasing effect of PBH in the case where a single source causes both observed CMB temperature perturbations and
PBH formation. In section~\ref{Different source case}, 
we discuss the other case where the source of PBH formation is different from that of CMB perturbations. 
Section~\ref{Conclusions} is devoted to the conclusions.

\section{Single source case}\label{Single source case}

In this section we discuss the simplest case that the fluctuation of a certain single field causes 
both observed large scale CMB temperature perturbations and PBH formation.
Let us denote a primordial adiabatic curvature perturbation on the comoving slice as ${\mathcal R}$. 
Although the amplitude of $\mathcal{R}$ is known to be around $10^{-5}$ on CMB scale,
it could be large enough to produce PBHs on smaller scales, becoming blue-tilted~\cite{Kohri:2007qn,Drees:2011hb} 
or having some bump on its power spectrum~\cite{Kawasaki:1997ju,Yokoyama:1998pt,Kawaguchi:2007fz,Frampton:2010sw}.
Also we assume $\mathcal{R}$ has a small local-type non-Gaussianity
which is represented as
\begin{eqnarray}\label{local-type non-Gaussianity}
	\mathcal{R}(\mathbf{x})=g(\mathbf{x})+f_\mathrm{NL}(g^2(\mathbf{x})-\braket{g^2}),
\end{eqnarray}
where $g$ is a Gaussian field. 
Here, we do not mention the concrete PBH formation model where the generated adiabatic curvature perturbation has the above type of non-Gaussianity.
There have been several claims about the local-type non-Gaussianity in single source case.
According to Maldacena~\cite{Maldacena:2002vr},
there is a consistency relation between non-linearity parameter $f_\mathrm{NL}$ and the spectral tilt of the adiabatic curvature perturbations,
and hence even in the simplest inflation model, namely single-field slow-roll inflation, the local-type non-Gaussianity does not vanish and is $O(10^{-2})$.
On the other hand,  recently, some authors claimed that such a non-Gaussianity given by the consistency relation is nothing but 
a gauge artifact and it does not contribute the scale-dependent bias~\cite{Tanaka:2011aj,Creminelli:2011sq,Pajer:2013ana} and this statement is based on
the adiabaticity of the curvature perturbations.
As for the PBH formation model, e.g., in the bump cases, in order to grow up the curvature perturbations rapidly on the bump scale
the adiabaticity would be violated and it would be possible to generate the curvature perturbations with  non-zero local-type non-Gaussianity.
Related issues in preheating model was discussed in Ref.~\cite{Bond:2009xx}.
Anyway we simply presume a small non-Gaussianity $f_\mathrm{NL}\sim\mathcal{O}(0.01)$ here without any assumption of the concrete inflationary models.

\subsection{PBH bias}
We will assume that PBH formation can be described by Press-Schechter approach~\cite{Press:1973iz} 
as in the standard structure formation.
Following the peak background split picture in the structure formation~\cite{Bardeen:1985tr},
we discuss the large scale clustering of PBH.
In other words, we will consider how the PBH number density in some region of scale $R$ will be modulated if that region is on some long-wavelength
fluctuations whose scale $R_l$ is much larger than $R$. Anyway let us calculate the PBH density without the long-wavelength mode at first.
We would like to note that using the peak background split picture is just for intuitive understanding how the non-Gaussianity affects the large scale clustering
of the biased objects such as halos, galaxies and PBHs.
A clustering behavior of  halos/galaxies with the non-Gaussian initial fluctuations has been also discussed by employing not only the peak-background split picture 
but also other analytic approach \cite{Matarrese:2008nc,Matsubara:2012nc}. There are also several literatures about numerical N-body simulation with the non-Gaussian initial fluctuations
and they are consistent with the analytic result (see, e.g., Ref. \cite{Dalal:2007cu}).

According to~\cite{Young:2014ana}, when the PBH formation is discussed, one should not use the curvature perturbations $\mathcal{R}$ but take 
the density perturbations of the radiation on the comoving slice $\delta$.
The Fourier transformed component of such density perturbations of the radiation is obtained from the Poisson equation and it becomes~\cite{Young:2014ana}
\begin{eqnarray}
	\delta(\mathbf{k})=\frac{4}{9}\left(\frac{k}{aH}\right)^2\mathcal{R}(\mathbf{k}),
\end{eqnarray}
on super-horizon scale. Following the Press-Schechter approach, we consider the density perturbations coarse-grained on 
PBH-formation scale. 
With some window function $W(kR_s)$, we define the coarse-grained density perturbations as
\begin{eqnarray}\label{def of ds}
	\delta_s(\mathbf{k})=\mathcal{M}_s(k)\mathcal{R}(\mathbf{k}),
\end{eqnarray}
where
\begin{eqnarray}\label{Ms}
	\hspace{-15pt}\mathcal{M}_s(k)\!=\!\frac{4}{9}\!\left(\!\frac{k}{aH|_\mathrm{PBH}}\!\right)^2\!W(kR_s)
	\!=\!\frac{4}{9}(kR_\mathrm{PBH})^{2}W(kR_s).
\end{eqnarray}
$R_s$ is the coarse-graining scale, and almost the same as the horizon scale at the PBH-formation time 
$R_\mathrm{PBH}=(aH|_\mathrm{PBH})^{-1}$ since
the overdensity will collapse soon after entering the horizon. Note that $R$ should be much larger than $R_s$.

In the Press-Schechter approach, the overdensity such that $\delta_s(\mathbf{x})$ exceeds a certain threshold $\delta_c$ is assumed to 
collapse to be PBHs.\footnote{In the case such that $\mathcal{R}$ has the non-Gaussianity, the dynamics of collapse after the 
overdensity enters the horizon may also vary. Therefore we may not be able to say sweepingly the overdensity such that $\delta_s>\delta_c$ will
collapse (For instance, Nakama \emph{et al.}~\cite{Nakama:2013ica} analyzed the development of overdensities of generalized curvature profile
beyond the Gaussian one and proposed two crucial parameters as thresholds of PBH formation.). 
However those effects are beyond the scope of this paper, and we just follow the standard Press-Schechter approach.}
Therefore, assuming the statistics of $\delta_s$ is also nearly Gaussian,\footnote{The effect of the non-Gaussian profile of $\delta_s$
is considered in~\cite{Saito:2008em,Byrnes:2012yx,Young:2014oea,Young:2015kda} for example. However such an effect to the scale-dependent bias is 
the second order quantity w.r.t. $f_\mathrm{NL}$~\cite{Desjacques:2008vf,Desjacques:2011jb,Yokoyama:2012az}, and therefore it can be neglected safely. 
Moreover, in appendix \ref{Chi-squared type}, it is suggested that the part of bias factor without $f_\mathrm{NL}$ is not so different 
even if the non-Gaussian profile effect is taken into account as long as the PBH abundance is fixed.}
the probability of PBH formation is given by
\begin{eqnarray}\label{P1}
	P_1(>\nu)&=&\frac{2}{\sqrt{2\pi\sigma_s^2}}\int^\infty_{\delta_c}\dd\delta\exp\left(-\frac{\delta^2}{2\sigma_s^2}\right) \nonumber \\
	&=&\mathrm{erfc}\left(\frac{\nu}{\sqrt{2}}\right)\simeq\sqrt{\frac{2}{\pi}}\frac{1}{\nu}\ee^{-\nu^2/2},
\end{eqnarray}
where $\nu=\delta_c/\sigma_s$ and $\sigma_s$ is the standard deviation of $\delta_s$ 
averaged in the region $R$:
\begin{eqnarray}\label{def of variance}
	\sigma_s^2&=&\braket{\delta_s^2}_R=\frac{1}{V}\int_R\dd^3x\,\delta_s^2(\mathbf{x}) \nonumber \\
	&=&\frac{1}{V}\int_R\dd^3x\int\frac{\dd^3k\,\dd^3q}{(2\pi)^6}\mathcal{M}_s(k)\mathcal{M}_s(q)\mathcal{R}(\mathbf{k})\mathcal{R}(\mathbf{q})\ee^{i(\mathbf{k}+\mathbf{q})\cdot\mathbf{x}}, \nonumber \\
\end{eqnarray}
where $V$ is the volume of the region $R$.
We have included conventional factor 2 of Press-Schechter formulation.
Also we have assumed a high peak limit as $\nu\gg1$, because PBHs should be rare objects.
Then we will consider how this probability will be modulated under the existence of long-wavelength modes.

\begin{figure}
	\includegraphics[width=0.9\hsize]{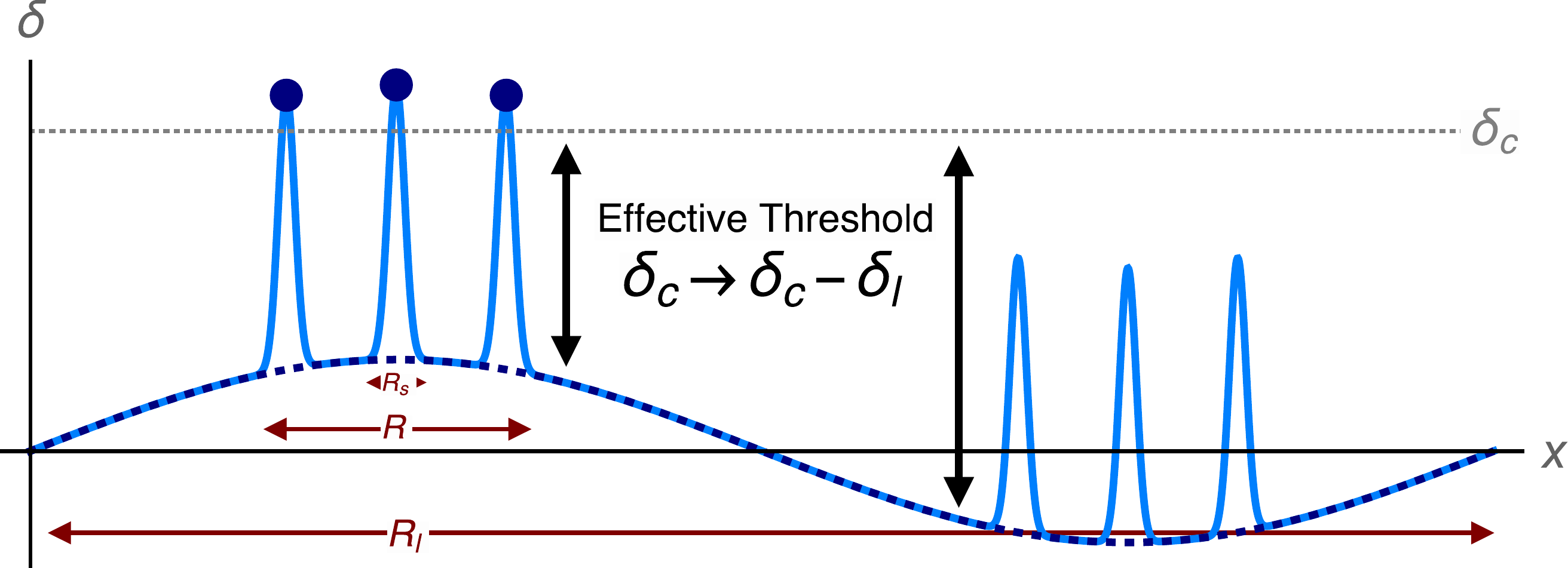}
	\\[10pt]
	\includegraphics[width=0.9\hsize]{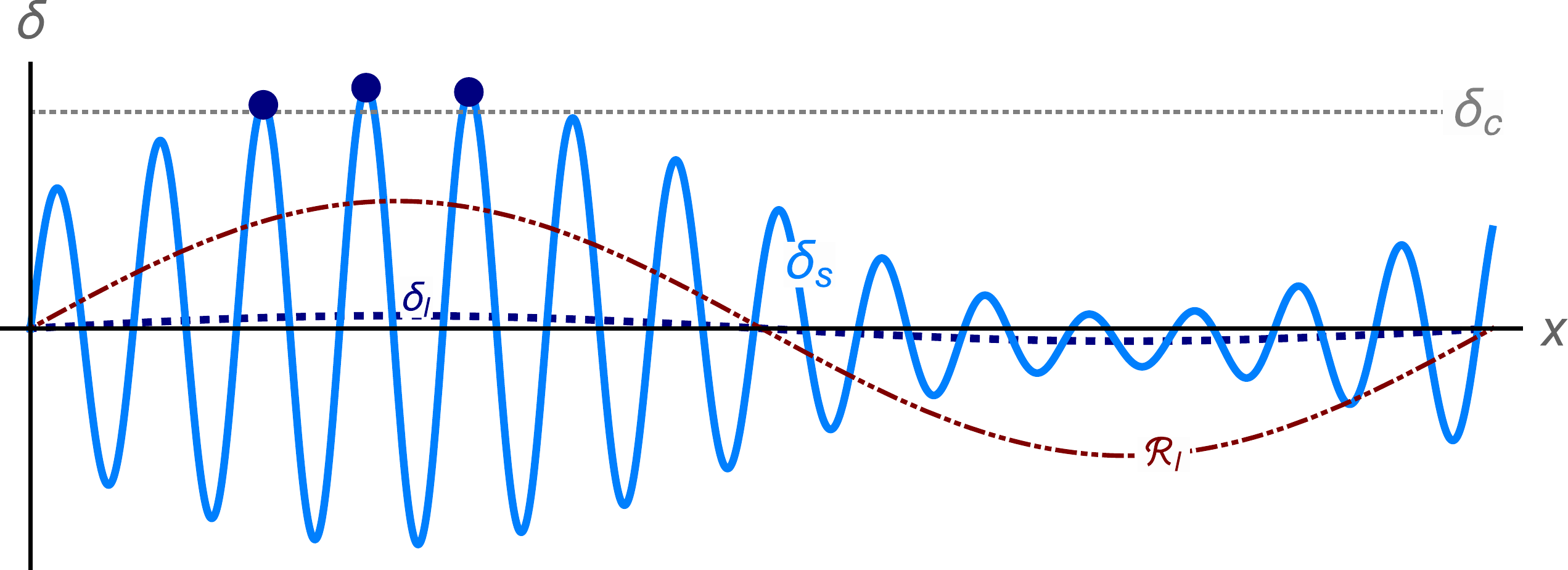}
	\caption{Schematic diagrams of a bias effect in the peak-background split formulation.
	Gray dotted lines represent the threshold for PBH formation $\delta_c$ and dark blue points show the regions which will be PBHs.
	In the first (top) figure, it can be seen that the threshold for the short-wavelength mode $\delta_s$ is effectively 
	reduced to $\delta_c-\delta_l$. Therefore PBHs tend to be formed in $\delta_l>0$ region, and this effect is called the scale-independent bias.
	However, long-wavelength modes of the comoving radiational perturbations are practically quite suppressed compared to 
	the curvature perturbation $\mathcal{R}_l$ 
	due to the factor $(kR_\mathrm{PBH})^2$
	in Eq.~(\ref{Ms}) as the second (bottom) figure shows. Even in such a case, if $f_\mathrm{NL}>0$ (or $<0$), 
	the amplitude of $\delta_s$ itself tends to be larger, being proportional direct to $\mathcal{R}_l$ via the non-Gaussianity of
	primordial curvature perturbations. Such a bias effect is called the scale-dependent bias.}
	\label{bias image}
\end{figure}

With the long-wavelength mode, the two quantities $\delta_c$ and $\sigma_s$ seem to be able to be modified.
However, in the integral of Eq.~(\ref{def of variance}), the factor $(k/aH)^2$ in $\mathcal{M}_s(k)$ cuts off the long-wavelength modes of $\mathcal{R}$
and $\sigma_s$ is determined only by the short-wavelength modes as
\begin{eqnarray}
	\hspace{-5pt}\sigma_s^2\!=\!\int_{k\!\gtsim\! R^{-1}}\!\dd\log k\,\mathcal{P}_\mathcal{R}(k)\mathcal{M}_s^2(k)\!\sim\!\mathcal{P}_\mathcal{R}(k_s)\mathcal{M}_s^2(k_s),
\end{eqnarray}
with the power spectrum $\mathcal{P}_\mathcal{R}(k)=\frac{k^3}{2\pi^2}\mathcal{R}(\mathbf{k})\mathcal{R}(-\mathbf{k})=\frac{k^3}{2\pi^2}|\mathcal{R}(\mathbf{k})|^2$.
Here we have replaced $V^{-1}\int_R\dd^3x\,\ee^{i(\mathbf{k}+\mathbf{q})\cdot\mathbf{x}}$ by $(2\pi)^3\delta^{(3)}(\mathbf{k}+\mathbf{q})$ for sufficiently large $k$ and $q$ compared to $R^{-1}$.
If $\mathcal{R}$ is completely Gaussian and there is no modal coupling between long- and short-wavelength modes, it does not affected by long modes at all.
On the other hand, the threshold $\delta_c$ is simply reduced to $\delta_c-\delta_l$ if the region $R$ is on the long-wavelength density perturbation $\delta_l$ (see Fig.~\ref{bias image}).
Therefore the PBH density in that region will be slightly larger as
\begin{eqnarray}
	\delta_\mathrm{PBH}(\mathbf{x})&:=&\frac{P_1(>\nu|\delta_l(\mathbf{x}))}{P_1(>\nu)}-1 \nonumber \\
	&\simeq&\pdif{\delta_c}{\delta_l}\pdif{\log P_1}{\delta_c}\delta_l(\mathbf{x})\simeq\frac{\nu}{\sigma_s}\delta_l(\mathbf{x}).
\end{eqnarray}
Here we have used a high peak limit $\nu\gg1$ again. The spatial label $\mathbf{x}$ denotes the coarse-grained position of the region $R$.
This coefficient $b_0:=-\pdif{\log P_c}{\delta_c}\simeq\frac{\nu}{\sigma_s}$ is called \emph{scale-independent bias}.

Thus, following the peak-background split picture, we find that
the number density fluctuations of PBH, $\delta_{\rm PBH}$, on large scales
is given by
 $\delta_\mathrm{PBH}\simeq\frac{\nu}{\sigma_s}\delta$ in the pure Gaussian case. However,
although a biasing factor $\nu / \sigma_s$ is larger than order of unity,
$\delta_{\rm PBH}$ is actually much smaller than $\mathcal{R}$ on the CMB scale.
This is because the CMB scale $k_\mathrm{CMB}^{-1}$ is quite larger than the PBH scale $R_\mathrm{PBH}$,
namely $k_\mathrm{CMB}R_\mathrm{PBH}\ll 1$, and then
$\delta_{\rm PBH}$ is strongly suppressed as $\delta_{\rm PBH}(k_\mathrm{CMB}) \sim \nu/\sigma_s \times 
(k_\mathrm{CMB}R_\mathrm{PBH})^2{\mathcal R}(k_\mathrm{CMB}) \ll {\mathcal R}(k_\mathrm{CMB}) $.
Indeed, this result is consistent with the claims of \cite{Young:2014ana,Chisholm:2005vm,Nakama:2014fra}.
However, it will be dramatically changed if we consider the effect of non-Gaussianity, namely the scale-dependent bias~\cite{Dalal:2007cu}. 

If the primordial perturbations have local-type non-Gaussianity as given in Eq.~(\ref{local-type non-Gaussianity}), there is a correlation between short- and long-wavelength perturbations as shown later.
Therefore the long-wavelength density perturbation $\delta_l$ not only reduces the threshold effectively but also modifies the variance of
short-wavelength perturbations, $\sigma_s$. Thus the bias parameter $b=\left.\dif{\log P_1}{\delta_l}\right|_{\delta_l=0}$ has the other component
\begin{eqnarray}
	\Delta b=\left.\pdif{\log\sigma_s}{\delta_l}\pdif{\log P_1}{\log\sigma_s}\right|_{\delta_l=0},
\end{eqnarray}
which is called \emph{scale-dependent bias}. Then let us evaluate this bias effect.

From Eq.~(\ref{local-type non-Gaussianity}),  $\mathcal{R}(\mathbf{x})$ can be decomposed into long- and short-wavelength components as
\begin{eqnarray}\label{decomposition into long and short wavelength mode}
	\mathcal{R}(\mathbf{x})&\!=\!&g_l(\mathbf{x})\!+\!g_s(\mathbf{x})\!+\!f_\mathrm{NL}[(g_l(\mathbf{x})\!+\!g_s(\mathbf{x}))^2\!-\!
	\braket{g_l^2}\!-\!\braket{g_s^2}] \nonumber \\
	&\!=\!&g_l(\mathbf{x})\!+\!f_\mathrm{NL}(g_l^2(\mathbf{x})\!-\!\braket{g_l^2}) \nonumber \\
	&&+g_s(\mathbf{x})\!+\!f_\mathrm{NL}(2g_l(\mathbf{x})g_s(\mathbf{x})\!+\!g_s^2(\mathbf{x})\!-\!\braket{g_s^2}).
\end{eqnarray}
In the second equation, the first line which is independent of $g_s$ denotes the long-wavelength mode $\mathcal{R}_l$ and
the second line is regarded as the short-wavelength mode $\mathcal{R}_s$. 
$\mathcal{R}_l$ will be cut off in the integral for $\sigma_s^2$~(\ref{def of variance}) due to the transfer function $\mathcal{M}_s(k)$ as we have mentioned,
but now we have another contribution of long-wavelength modes to the variance via $g_l(\mathbf{x})$ in $\mathcal{R}_s$.
Since the second order terms of $g_s(\mathbf{x})$ do not 
include the long-wavelength mode $g_l(\mathbf{x})$ and their contributions to bias effects are suppressed as the second order of $f_\mathrm{NL}$~\cite{Desjacques:2008vf,Desjacques:2011jb,Yokoyama:2012az},
one can neglect these terms and 
see that the variance $\sigma_s^2$
in the region $R$ is just constantly amplified as
\begin{eqnarray}
	\sigma_s(\mathbf{x})=(1+2f_\mathrm{NL}g_l(\mathbf{x}))\bar{\sigma}_s,
\end{eqnarray}
noting that $g_l(\mathbf{x})$ is almost constant in the region $R$.
Here $\bar{\sigma}_s^2$ denotes the variance of $\delta_s$ averaged over the whole universe, namely $\bar{\sigma}_s^2=\braket{\delta_s^2}$.
Hence, with an approximation $\mathcal{R}_l(\mathbf{x})\simeq g_l(\mathbf{x})$, the Fourier mode of the dependence of $\sigma_s$ to $\delta_l$ on large scale is 
given by
\begin{eqnarray}
	\left.\pdif{\log\sigma_s(k)}{\delta_l(k)}\right|_{\delta_l(k)=0}&\simeq&\left.\mathcal{M}_l^{-1}(k)\pdif{\log\sigma_s(k)}{g_l(k)}\right|_{g_l(k)=0}
	\nonumber \\
	&=&2f_\mathrm{NL}\mathcal{M}_l^{-1}(k),
\end{eqnarray}
where, 
\begin{eqnarray}
	\mathcal{M}_l(k)=\frac{4}{9}(kR_\mathrm{PBH})^2W(kR_l).
\end{eqnarray}
On the other hand, since $P_1$ is the function only of $\nu$, we can obtain the following relation with use of the definition of $b_0$.
\begin{eqnarray}\label{dlogP1/dlogsigmas}
	\hspace{-5pt}\pdif{\log P_1}{\log\sigma_s}=\sigma_s\dif{\nu}{\sigma_s}\dif{\log P_1}{\nu}
	=-\delta_c\pdif{\log P_1}{\delta_c}=\delta_cb_0.
\end{eqnarray}
From the above results, the scale-dependent bias is given by
\begin{eqnarray}
	\Delta b(k)=2f_\mathrm{NL}\mathcal{M}_l^{-1}(k)\delta_cb_0.
	\label{scaledepbias}
\end{eqnarray}	
It is noteworthy that $\Delta b$ has a factor of $\mathcal{M}_l^{-1}$. Since this factor cancels out $(kR_\mathrm{PBH})^2$
in $\delta_s$,
$\delta_\mathrm{PBH}$ with the scale-dependent bias is not negligible compared to the adiabatic curvature perturbation ${\mathcal R}$ 
even on much larger scales.
This is because the scale-dependent bias is directly proportional to $\mathcal{R}$ via the non-Gaussianity, while
the scale-independent bias is proportional to $\delta$. 	

\subsection{PBH-DM isocurvature perturbation}
Gravitational collapse is irreversible process and potentially accompanied by entropy production.
Furthermore, especially in the case of PBH formation, the equation of state of the density component changes from the radiation to the matter,
and therefore the density perturbations of PBHs can bring about easily-detectable entropic (or isocurvature) perturbations.
In this section we will calculate the specific value of the isocurvature perturbations.

If a dominant component of DMs consists of PBH, the matter isocurvature perturbation can be given in terms of $\delta_{\rm PBH}$ as
\begin{eqnarray}
	S: =\delta_\mathrm{PBH}-\frac{3}{4}\delta.
\end{eqnarray}
By using  Eq. (\ref{scaledepbias}), the Fourier transformed component of the matter isocurvature perturbation 
smoothed on the scale $R$ is
\begin{eqnarray}	
	S_{R}(\mathbf{k})&\!=\!&\left(b_0\!+\!\Delta b(k)\!-\!\frac{3}{4}\right)\delta_{R}(\mathbf{k}) \nonumber \\
	&\!=\!&\!\left(\!b_0\mathcal{M}_{R}(k)\!+\!2f_\mathrm{NL}\delta_cb_0\!\frac{\mathcal{M}_{R}}{\mathcal{M}_l}\!(k)
	\!-\!\frac{3}{4}\mathcal{M}_{R}(k)\!\right)\!\mathcal{R}(\mathbf{k}), \nonumber \\
\end{eqnarray}
where $\mathcal{M}_R(k)=\frac{4}{9}(kR_\mathrm{PBH})^2W(kR)$.
As mentioned before, the factor $(k_\mathrm{CMB}R_\mathrm{PBH})^2$ in $\mathcal{M}_{R}$ 
is so small that we can neglect the first and third terms in the parenthesis. 
Thus the power spectrum of the isocurvature mode is given by\footnote{Note that this result does not mean
the rarer PBHs bring about the more isocurvature perturbations even though the higher $\nu$ means the lower abundance of PBHs.
This result supposes $\Omega_\mathrm{PBH}=\Omega_\mathrm{DM}$, and if PBHs are sub-component of DMs, 
the isocurvature modes are proportional to $\left(\frac{\Omega_\mathrm{PBH}}{\Omega_\mathrm{DM}}\right)^2$ which is in fact exponentially suppressed
as shown in Eq.~(\ref{P1}).}
\begin{eqnarray}\label{power of S}
	\mathcal{P}_S(k_\mathrm{CMB})&\simeq&(2f_\mathrm{NL}\delta_cb_0)^2\mathcal{P}_\mathcal{R}(k_\mathrm{CMB}) \nonumber \\
	&\simeq&\left(2f_\mathrm{NL}\nu^2\right)^2\mathcal{P}_\mathcal{R}(k_\mathrm{CMB}).
\end{eqnarray}
Here we have approximated $\frac{\mathcal{M}_{R}}{\mathcal{M}_l}(k_\mathrm{CMB})$ as unity because $k_\mathrm{CMB}^{-1}$ is
larger than both $R_l$ and $R$ now and then $W(k_\mathrm{CMB}R_l)\simeq W(k_\mathrm{CMB}R)\simeq1$.
Note that $S_R$ is similarly equal to the unsmoothed mode $S$ on CMB scale. Also
we have used $\nu=\delta_c/\sigma_s$ and $b_0\simeq\nu/\sigma_s$ in the second equation.

CMB observations have already given tight constraints on the amplitude of the matter isocurvature perturbations
and,
according to the Planck collaboration~\cite{Ade:2013uln}, 
in the case of fully (anti-) correlated type which corresponds to $S\propto\mathcal{R}$ (or $S\propto-\mathcal{R}$),
the matter isocurvature mode is constrained as
\begin{eqnarray}\label{original isocurvature constraint}
	\hspace{-5pt}\frac{\mathcal{P}_S}{\mathcal{P}_S\!+\!\mathcal{P}_\mathcal{R}}\!\ltsim\!
	\begin{cases}
		0.0025 & \text{(fully correlated, $f_\mathrm{NL}\!>\!0$)} \\
		0.0087 & \text{(anti-correlated, $f_\mathrm{NL}\!<\!0$)},
	\end{cases}
\end{eqnarray}
at $95\%$ CL.
Therefore, from Eq. (\ref{power of S}) we must satisfy
\begin{eqnarray}\label{isocurvature constraint}
	|f_\mathrm{NL}|\nu^2\ltsim
	\begin{cases}
		\displaystyle
		\frac{\sqrt{0.0025}}{2}=0.025. & (f_\mathrm{NL}>0) \\[5pt]
		\displaystyle
		\frac{\sqrt{0.0087}}{2}=0.047. & (f_\mathrm{NL}<0)
	\end{cases}
\end{eqnarray}
Assuming the nearly monochromatic mass function for PBHs, 
the current abundance of PBHs is given by~\cite{Carr:2009jm}
\begin{eqnarray}
	\Omega_\mathrm{PBH}\sim0.86\times10^8P_1\left(\frac{M}{M_\odot}\right)^{-1/2},
\end{eqnarray}
where $M_\odot$ denotes the solar mass $\sim2\times10^{33}\,\mathrm{g}$. Inversely, if DMs consist of PBHs and $\Omega_\mathrm{PBH}
=\Omega_\mathrm{DM}=0.31$~\cite{Ade:2013zuv}, the probability $P_1$ should satisfy the following relation
\begin{eqnarray}
	P_1\sim0.36\times10^{-8}\left(\frac{M}{M_\odot}\right)^{1/2}.
\end{eqnarray}
Furthermore, from Eq.~(\ref{P1}), $\nu$ can be written in terms of $P_1$ as
\begin{eqnarray}
 	\nu=\sqrt{2}\,\mathrm{erfc}^{-1}(P_1),
\end{eqnarray} 
where $\mathrm{erfc}^{-1}$
denotes the inverse function of the complementary error function.
Therefore, if we assume the mass of PBH is less than $10^{40}\,\mathrm{g}$,\footnote{
Here we respect the plot of Ref.~\cite{Capela:2013yf} in which the PBH constraints were summarized up to 
$M_\mathrm{PBH}=10^{40}\,\mathrm{g}$.
Moreover Afshordi \emph{et al.} showed that for more massive PBHs than $\sim10^{37}\,\mathrm{g}$, the Poisson-noise fluctuations deviating
from the adiabatic perturbations become too large to be consistent with observed $\sigma_8$~\cite{Afshordi:2003zb}. 
Therefore we constrain the PBH mass to
be less than around these values. Also, lighter PBHs than $10^{15}\,\mathrm{g}$ have been already evaporated due to the Hawking radiation by today, and
therefore we are interested only in more massive PBHs than $10^{15}\,\mathrm{g}$} 
we obtain the following relation
\begin{eqnarray}\label{lower bound for nu2}
	\hspace{-5pt}\nu^2
	\!\gtsim\!2\left[\mathrm{erfc}^{-1}\!\left(0.36\!\times\!10^{-8}\!\left(\!\frac{10^{40}}{2\!\times\!10^{33}}\!\right)^{1/2}\,\!\right)\!\right]^2
	\!\simeq\!20.
\end{eqnarray}
Here, note that $\mathrm{erfc}^{-1}(x)$ is a monotonic decreasing function.
It obviously conflicts with the constraint~(\ref{isocurvature constraint}) with $f_\mathrm{NL}\gtsim\mathcal{O}(0.01)$.

As a result, we find that it is difficult to satisfy the isocurvature constraint~(\ref{isocurvature constraint}) even with a small non-linearity 
parameter $|f_\mathrm{NL}| \sim\mathcal{O}(0.01)$ and this means that the PBH-DM scenario is strictly constrained
due to the existence of the small local-type non-Gaussianity
in the single sourced case.
This is a main result of this paper. 

\begin{figure}
	\begin{center}
		\includegraphics[width=0.9\hsize]{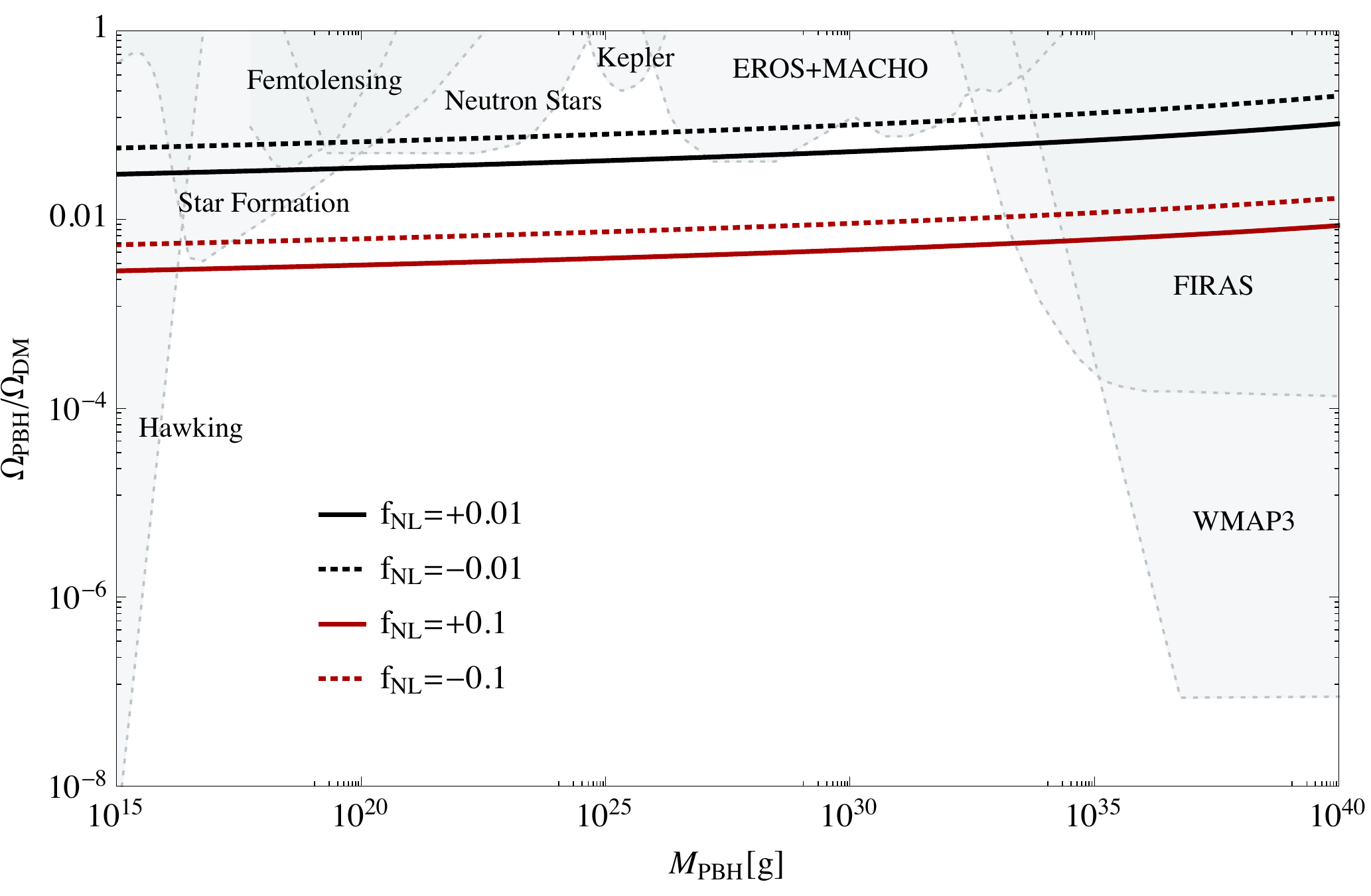}
		\caption{The plot of the upper limit of the current PBH fraction $\Omega_\mathrm{PBH}/\Omega_\mathrm{DM}$ from the 
		isocurvature constraints~(\ref{original isocurvature constraint}). Shaded regions represent existing constraints from various 
		observations~\cite{Griest:2013aaa,Capela:2013yf}.
		The black solid and dotted lines show the upper limit in the case of $f_\mathrm{NL}=\pm0.01$ and 
		the red solid and dotted ones exhibit the constraints in the case of $f_\mathrm{NL}=\pm0.1$.
		Also from this plot, it can be seen that PBHs cannot be the main component of DMs even with small non-linearity parameter 
		$|f_\mathrm{NL}|=0.01$.}
		\label{PBHconstraints_final}
	\end{center}
\end{figure}

Furthermore, we have also considered the case that PBH is a subdominant component of DMs.
We have plotted the upper limit of the PBH fraction in such a case
to satisfy the isocurvature constraints~(\ref{original isocurvature constraint}) in Fig.~\ref{PBHconstraints_final}.
Note that the isocurvature power and the probability of PBH formation are 
given by $\mathcal{P}_S=\left(\frac{\Omega_\mathrm{PBH}}{\Omega_\mathrm{DM}}\right)^2
\mathcal{P}_\mathrm{PBH}$ and $P_1\sim0.36\times10^{-8}\frac{\Omega_\mathrm{PBH}}{\Omega_\mathrm{DM}}\left(\frac{M}{M_\odot}\right)^{1/2}$
respectively in that case.

\section{Different source case}\label{Different source case}
In the case where the source fluctuation of PBH formation is different from that of CMB fluctuations, the above discussion does not change so much.
Let $\mathcal{R}$ and $\Sigma$ denote the curvature perturbations which respectively induce  the CMB fluctuations and the PBH formation.
In other words, $\mathcal{R}$ is dominant and has a nearly flat spectrum on CMB scale, while $\Sigma$ dominates on much smaller scale.

If we assume $\Sigma$ is almost Gaussian but has a small local-type non-Gaussianity as
\begin{eqnarray}
	\Sigma(\mathbf{x})=g(\mathbf{x})+f_{\mathrm{NL},\Sigma}(g^2(\mathbf{x})-\braket{g^2}),
\end{eqnarray}
the power of the matter isocurvature perturbations is given by 
\begin{eqnarray}
\mathcal{P}_S\simeq\left(2f_{\mathrm{NL},\Sigma}\nu^2\right)^2\mathcal{P}_\Sigma,
\end{eqnarray}
like as Eq.~(\ref{power of S}).
By introducing an effective non-linearity parameter defined as
\begin{eqnarray}
f_\mathrm{NL,eff} := f_{\mathrm{NL},\Sigma}\sqrt{\frac{\mathcal{P}_\Sigma}{\mathcal{P}_\mathcal{R}}},
\end{eqnarray}
the power of the isocurvature perturbation can be written as 
\begin{eqnarray}
\mathcal{P}_S=\left(2f_{\mathrm{NL,eff}}\nu^2\right)^2\mathcal{P}_\mathcal{R}.
\end{eqnarray}
From the above expression, 
we can also obtain the constraints on the PBH-DM scenario in the different source case depending
on the ratio between the power of ${\mathcal R}$ and that of $\Sigma$ on CMB scale.
Note that in this case the correlation between adiabatic curvature perturbations and generated isocurvature perturbations 
is much suppressed, that is, almost uncorrelated-type  isocurvature perturbations.
The constraint for the uncorrelated isocurvature perturbations shown by Planck collaboration is~\cite{Ade:2013uln}
\begin{eqnarray}
	\frac{\mathcal{P}_S}{\mathcal{P}_S+\mathcal{P}_\mathcal{R}}\ltsim0.036, \quad \text{(uncorrelated)},
\end{eqnarray}
at $95\%$ CL.
and therefore the constraint for $f_\mathrm{NL,eff}$ is weakened up to 
\begin{eqnarray}
	|f_\mathrm{NL,eff}| \,\nu^2\ltsim0.095.
\end{eqnarray}
Furthermore, $\Sigma$ is not necessarily almost Gaussian but possibly quite non-Gaussian. Accordingly, we also discuss the
chi-squared case in appendix~\ref{Chi-squared type} as a simple example.

\section{Conclusions}\label{Conclusions}	
In this paper, we consider the clustering of the PBHs on CMB scale produced by the bias effect.
Since there is a great gap between CMB scale and the horizon scale at the time of PBH formation during the radiation dominant phase,
the distribution of PBH is hardly biased basically. However, if the curvature perturbation field which cause PBHs 
has a local-type non-Gaussianity, even if it is small, we have showed non-negligible spatial fluctuations of PBH number density
can be produced on CMB scale 
because the source field has a correlation between large and small scale via the non-Gaussianity.	
Specifically, the fluctuation of the number density of PBH is biased from the source curvature perturbation roughly by the factor of non-linearity 
parameter $f_\mathrm{NL}$
times $\left(\mathrm{erfc}^{-1}(P_1)\right)^2$, where $P_1$ is the PBH production probability which is commonly represented as $\beta$ in
the literature. If DMs consist of PBHs, these extra density perturbations on CMB scale can be detected as the matter isocurvature perturbation
and are constrained strictly by CMB observations. 

We have showed, if the source curvature perturbation of PBHs also cause
CMB temperature fluctuations, the PBH-DM scenario is excluded even with quite small non-linearity parameter 
$|f_\mathrm{NL}| \sim\mathcal{O}(0.01)$.
On the other hand, in the case that 
the source curvature perturbation of PBHs $\Sigma$ is different from that of CMB fluctuations $\mathcal{R}$,
we can avoid the isocurvature constraints if $\Sigma$ is sub-dominant enough on CMB scale. Even such a case,
the PBH-DM scenario should satisfy $|f_{\mathrm{NL},\Sigma}|\sqrt{\frac{\mathcal{P}_\Sigma}
{\mathcal{P}_\mathcal{R}}(k_\mathrm{CMB})}\ltsim\mathcal{O}(0.001)$.
These constraints are almost independent of the PBH mass, and moreover they may not change a lot even if the PBH-mass spectrum
is quite non-monochromatic unlike many other constraints. This is because even if the isocurvature perturbations of each mass range are 
small enough, they will be piled up and the total amplitude of them will be determined only by $f_\mathrm{NL}$ and $\Omega_\mathrm{PBH}$, hardly
affected by the shape of mass function. We leave more precise discussion about the effect of non-monochromaticity for a future issue.

We also  think that it should be interesting to evaluate the specific value of $f_\mathrm{NL}$ in some concrete cases where abundant PBHs are produced.
Further, though we focused on simple local-type non-Gaussianity here, it will be interesting and important to discuss the bias effect
for other types of non-Gaussianity. We also leave such studies for 
future issues.

\begin{acknowledgements}
We would like to thank Masahiro Kawasaki and Tomohiro Harada for useful comments.
This work was supported by World Premier International Research Center Initiative (WPI Initiative), MEXT, Japan. 
YT is supported by an Advanced Leading Graduate Course for Photon Science grant.
SY acknowledges the support by Grant-in-Aid for JSPS Fellows No. 242775.
\end{acknowledgements}

\appendix

\section{Chi-squared type}\label{Chi-squared type}
In this appendix, we consider the simple chi-squared $\Sigma$
which is characterized as
\begin{eqnarray}\label{chi-square}
	\Sigma(\mathbf{x})=g^2(\mathbf{x})-\braket{g^2},
\end{eqnarray}
as an example of fully non-Gaussian case.
Suppose the coarse-grained density perturbations $\delta_s$ also follow the chi-squared distribution, the probability of PBH formation
is given by
\begin{eqnarray}\label{P1 for chi}
	P_1(>\delta_c)&\!=\!&2 \int^\infty_{\delta_c}\dd\delta\frac{1}{\sqrt{\pi\sigma_s(\sqrt{2}\delta\!+\!\sigma_s)}}
	\exp\left(-\frac{\sqrt{2}\delta\!+\!\sigma_s}{2\sigma_s}\right) \nonumber \\
	&\!=\!& 2 \, \mathrm{erfc}\left(\sqrt{\frac{\sqrt{2}\nu+1}{2}}\right) \nonumber \\
	&\!\simeq\!& 2 \sqrt{\frac{2}{\pi(\sqrt{2}\nu+1)}}\exp\left(-\frac{\sqrt{2}\nu+1}{2}\right),
\end{eqnarray}
where $\nu=\delta_c/\sigma_s$ again.
The scale-independent bias can be calculated similarly to section~\ref{Single source case} as
\begin{eqnarray}
	b_0=-\pdif{\log P_1}{\delta_c}=-\frac{1}{\sigma_s}\dif{\log P_1}{\nu}\simeq\frac{1}{\sqrt{2}\sigma_s}.
\end{eqnarray}
On the other hand, to calculate the scale-dependent part of the bias parameter, we have to discuss the dependence of $\sigma_s$ on $\delta_l$.
From Eq.~(\ref{chi-square}), we decompose $\Sigma$ into long- and short-wavelength components like as 
Eq.~(\ref{decomposition into long and short wavelength mode}),
\begin{eqnarray}
	\Sigma(\mathbf{x})&=&g_l^2(\mathbf{x})-\braket{g_l^2} \nonumber \\
	&&+g_s^2(\mathbf{x})-\braket{g_s^2}+2g_l(\mathbf{x})g_s(\mathbf{x}).
\end{eqnarray}
Similarly to the discussion in section~\ref{Single source case}, the second line terms in the expression of $\Sigma (\mathbf{x})$
denote the short-wavelength mode of $\Sigma$.
Noting that third moments of $g_s$ should vanish, the power spectrum of $\delta_s$ on $R$-scale
is given by
\begin{eqnarray}
	\left.\mathcal{P}_{\delta_s}(k)\right|_{\Sigma_l(\mathbf{x})}&\!=\!&\mathcal{M}_s^2(k)[4(g_l^2(\mathbf{x})\!-\!\braket{g_l^2})\mathcal{P}_{g_s}(k)
	\!+\!\mathcal{P}_{\Sigma_s}(k)|_{\Sigma_l(\mathbf{x})=0}] \nonumber \\
	&\!=\!&\mathcal{M}_s^2(k)[4\Sigma_l(\mathbf{x})\mathcal{P}_{g_s}(k)\!+\!\mathcal{P}_{\Sigma_s}(k)|_{\Sigma_l(\mathbf{x})=0}].
\end{eqnarray} 
Then the standard deviation of $\delta_s$, which is the square root of the integration of $\mathcal{P}_{\delta_s}$, can be approximated by
\begin{eqnarray}
	\sigma_s(\mathbf{x})\simeq\sigma_s|_{\Sigma_l(\mathbf{x})=0}\left(1+\frac{2\sigma_{g,s}^2}{\sigma_s^2}\Sigma_l(\mathbf{x})\right),
\end{eqnarray}
where
\begin{eqnarray}
	\sigma_{g,s}^2=\int\dd\log k\,\mathcal{P}_g(k)\mathcal{M}_s^2(k).
\end{eqnarray}
Therefore
\begin{eqnarray}
	\left.\pdif{\log\sigma_s(k)}{\delta_l(k)}\right|_{\delta_l=0}=\frac{2\sigma_{g,s}^2}{\sigma_s^2}\mathcal{M}_l^{-1}(k),
\end{eqnarray}
and with use of Eq.~(\ref{dlogP1/dlogsigmas}), we obtain the scale-dependent bias as
\begin{eqnarray}
	\Delta b=\frac{2\sigma_{g,s}^2}{\sigma_s^2}\mathcal{M}_l^{-1}\delta_cb_0\simeq\frac{\sqrt{2}\sigma_{g,s}^2}{\sigma_s^2}
	\mathcal{M}_l^{-1}\nu.
\end{eqnarray}
Since it has the factor of $\mathcal{M}_l^{-1}$ again, the number density fluctuation of the PBH can be produced through this scale-dependent bias
even on CMB scale
and the power of matter isocurvature is given by
\begin{eqnarray}
	\mathcal{P}_S(k_\mathrm{CMB})\simeq\left(\frac{\sqrt{2}\sigma_{g,s}^2}{\sigma_s^2}\nu\right)^2\mathcal{P}_\Sigma(k_\mathrm{CMB}).
\end{eqnarray} 
Introducing the effective non-linearity parameter as $f_\mathrm{NL,eff}=\frac{\sqrt{2}{\sigma_{g,s}^2}}{\sigma_s^2}
\sqrt{\frac{\mathcal{P}_\Sigma}{\mathcal{P}_\mathcal{R}}(k_\mathrm{CMB})}$, we obtain the constraint as
\begin{eqnarray}
	f_\mathrm{NL,eff}\nu\ltsim0.095.
\end{eqnarray}
Note that $\nu$ is not squared unlike the case of small non-Gaussianity and this fact just comes from
the apparent difference of $P_1$, Eq.~(\ref{P1}) and (\ref{P1 for chi}).
The lower bound for $\nu$ given by
\begin{eqnarray}
	\nu=\frac{2\left[\mathrm{erfc}^{-1}\left(P_1/2\right)\right]^2-1}{\sqrt{2}}\gtsim 14,
\end{eqnarray}
is not so different from that of $\nu^2$ in the case of local-type non-Gaussian, Eq.~(\ref{lower bound for nu2}).

We can not calculate $f_\mathrm{NL,eff}$ without determining the detail profile of $\Sigma$ (or $g$) like a spectrum index,
but can understand this result qualitatively. First, let us approximate $\frac{\sigma_{g,s}^2}{\sigma_s^2}$ by 
$\frac{\mathcal{P}_g}{\mathcal{P}_\Sigma}(R_s^{-1})$.
In fact, the contribution around smoothing scale $R_s$ is dominant for the integrations for $\sigma_{g,s}^2$ and $\sigma_s^2$
due to the factor of $\mathcal{M}_s$. 
Moreover $\mathcal{P}_\Sigma$ can be approximated by $\mathcal{P}_g^2$. Therefore $\sigma_{g,s}^2/\sigma_s^2$ is roughly given by $\mathcal{P}_g^{-1}$. On the other hand, in the case of nearly scale-independent $g$, local non-linearity parameter $f_{\mathrm{NL},\Sigma}$
for $\Sigma$
can be written as 
$B_\Sigma(k_1,k_2,k_3)/(P_\Sigma(k_1)P_\Sigma(k_2)+\text{2 perms.})$, where $(2\pi)^3\delta^{(3)}(\mathbf{k}_1+\mathbf{k}_2+\mathbf{k}_3)
B_\Sigma(k_1,k_2,k_3)=\braket{\Sigma(\mathbf{k}_1)\Sigma(\mathbf{k}_2)\Sigma(\mathbf{k}_3)}$ and $P_\Sigma(k)=|\Sigma(\mathbf{k})|^2$.
According to Wick's theorem, we can obtain $B_\Sigma(k_1,k_2,k_3)\sim\mathcal{P}_g(P_g(k_1)P_g(k_2)+\text{2 perms.})$.
Moreover
$(P_g(k_1)P_g(k_2)+\text{2 perms.})/(P_\Sigma(k_1)P_\Sigma(k_2)+\text{2 perms.})\sim\mathcal{P}_g^{-2}$ if $g$ is almost scale-independent.
Therefore $\sigma_{g,s}^2/\sigma_s^2\sim\mathcal{P}_g^{-1}$ indeed 
denotes the effective local non-linearity parameter $f_{\mathrm{NL},\Sigma}$ in the case of chi-squared field.

\end{document}